\documentclass[aps,prl,twocolumn,groupedaddress,showpacs,showkeys]{revtex4}

\usepackage{graphicx}

\begin{document}

\newcommand{\be}{\begin{equation}}
\newcommand{\ee}{\end{equation}}

\title{Apparatus for measuring the thermal Casimir force at large distances.}

\date{\today}

\author{Giuseppe Bimonte}
\affiliation{Dipartimento di Scienze
Fisiche, Universit{\`a} di Napoli Federico II, Complesso Universitario
MSA, Via Cintia, I-80126 Napoli, Italy}
\affiliation{INFN Sezione di
Napoli, I-80126 Napoli, Italy }

\begin{abstract}

We describe a  Casimir apparatus based on a differential force measurement between a Au-coated sphere  and a planar slab divided in two regions, one of which is made of  high-resistivity (dielectric) Si, and the other of Au. The crucial feature of the setup  is a semi-transparent plane parallel conducting over-layer, covering both regions. 
The setup offers two important advantages over existing Casimir setups. On one hand it leads to a large amplification of the difference between the Drude and the plasma prescriptions that are currently used to compute the thermal Casimir force. On the other hand, thanks to the screening power of the over-layer,  it is in principle  immune from  electrostatic forces caused by potential patches on the plates surfaces, that plague  present large distance Casimir experiments.  If a  semi-transparent conductive over-layer with identical  patch structure over the Au-Si regions of the plate can be manufactured, similar to the opaque over-layers used in recent searches of non-newtonian gravitational forces based on the isoelectronic technique, the way will be paved for  a clear observation of the thermal Casimir force up to separations of several microns, and an unambiguous discrimination between the Drude and the plasma  prescriptions.

\end{abstract}

\maketitle

According to Quantum Theory, empty space is filled with  zero-point fluctuations of the electromagnetic (em) field.  Polarizable (but otherwise neutral) bodies modify the spectrum of these quantum fluctuations, and this results in a tiny force between two bodies in vacuum,  called Casimir force after the dutch physicist who discovered this phenomenon  \cite{Casimir48}. By carefully adding up the zero-point energies of the em modes between two perfectly conducting  infinite plane-parallel discharged mirrors at distance $a$ in vacuum, Casimir found that the mirrors attract each other with a (unit area) force of magnitude:
\be
F_C=\frac{\pi^2 \hbar c}{240\,a^4} \;.
\ee  
By modern experimental techniques the Casimir force between two metallic plates has been precisely measured in the last 20 years \cite{lamor1,umar,gianni,decca4,decca5,decca6,chang2}, and
several experiments have now observed the Casimir force for  surfaces made of  diverse materials like   semiconductors \cite{semic},   conductive oxides \cite{ito}, ferromagnetic materials \cite{bani2}, and in  liquids \cite{liq}. Superconducting Casimir devices have been studied as well \cite{ala1,ala2,super1,superc}.   The possible exploitation of the Casimir force in the actuation of  micro and nano machined devices,  recently spurred a lot of interest in  the Casimir effect for microstructured surfaces \cite{chen,chan,bao,chiu,bani,deccanat}. For  reviews see \cite{book1,book2}.

In his pioneering paper, Casimir considered two mirrors at zero temperature. At finite temperature, thermal fluctuations of the em field provide an extra contribution to the Casimir force, called {\it thermal} Casimir force. When considering the thermal Casimir force, an important length scale is the thermal length $\lambda_T=
\hbar c/(2 \pi k_B T)$ (for $T=300$ K $\lambda_{\rm 300 K}=1.2\; \mu$m).  For separations $a \lesssim \lambda_T$  the thermal force is much smaller than the zero-point force, but for $ a \gg  \lambda_T$  it represents the dominant contribution. Surprisingly enough, estimating  the precise magnitude of the thermal force for conducting surfaces turned out to be much more subtle than expected, and this is still an open problem as we write.
In essence, theoreticians cannot reach an agreement on how to estimate the contribution  of the zero-frequency transverse-electrical (TE $\omega=0$) mode, which to a large extent determines the magnitude of the thermal Casimir force. Two prescriptions have emerged: the so-called Drude prescription and the plasma prescription. Within the Drude prescription, the TE $\omega=0$ is computed in accordance with the Drude model for the permittivity of ohmic conductors $\epsilon_D(\omega)=1-\omega_p^2/\omega(\omega+{\rm i} \gamma)$ (where $\omega_p$ is the plasma frequency, and $\gamma$ the relaxation frequency). With this prescription, the TE $\omega=0$ mode contributes nothing. The plasma prescription, on the other hand, posits that the TE $\omega=0$ mode should be computed using the plasma model of IR optics $\epsilon_D(\omega)=1-\omega_p^2/\omega^2$,
and then one finds that the TE $\omega=0$ mode gives a  non-vanishing contribution.  
It has been shown that, in addition to predicting different magnitudes for the thermal force,  the two prescriptions have important thermodynamic consequences: while the Drude prescription leads to a violation of Nernst heat theorem (in the idealized case of two conducting plates with a perfect crystal structure) \cite{bezerra,bezerra2}, the plasma prescription violates the Bohr-van Leeuwen theorem of classical statistical physics \cite{Martin,bimo2}. 

The experimental situation is still far from clear. Several small distance experiments \cite{bani2,decca4,decca5,decca6,chang2}, probing separations below one $\mu$m,  appear to be in agreement with the plasma model, and to rule out the larger thermal force predicted by the Drude model. These experiments provide the most precise measurements of the Casimir force to date, with errors in the percent range.  One has to bear in mind however that in the submicron range the thermal force only represents a small correction to the zero-point force, and therefore its observation is very challenging. For a review of these experiments
see \cite{critiz}.  

In principle,   observing the thermal force  should be easier for $a \gtrsim \lambda_T$. 
For these large separations,  the thermal force is dominant and it should be easier to discriminate between the Drude and the plasma prescriptions, because they predict marketly different results. For example for two plane-parallel conducting surfaces, both models predict that for large separations the unit-area Casimir force attains a {\it universal} limit,  of magnitude $\zeta(3) k_B T /8 \pi a^3$ for the Drude model, and {\it twice} as large for the plasma model. Unfortunately, observation of the thermal force for separations in the micron region is very difficult too, because of the unavoidable presence of large electrostatic forces, that cannot be eliminated by applying a bias potential.  These forces originate from regions, called patches, of varying potential on the surfaces caused by spatial changes of crystalline structure and/or by adsorbed impurities. The force caused by these patches has been observed for example in an experiment with Al surfaces in \cite{antonini}, and it was found to be over 100 times   larger than the sought for thermal Casimir force,  in the range from 3.5 to 5 $\mu$m.  Large electrostatic forces were reported as well in a recent experiment by the Yale group \cite{lamorth}, which claims to have observed the thermal force between a large sphere and a plate both covered with gold, in the wide range of separations form 0.7 to 7.3 $\mu$m. The results have been interpreted by the authors as being in accordance with the  Drude prescription. This experiment has been criticized  \cite{critiz}, because the thermal Casimir force was obtained only after subtracting from the total measured force the much larger electrostatic force.
The subtraction was perfomed by making a fit of the total observed force, based on a two-parameter model of the electrostatic force,   and not by a direct and independent measurement, as it would have been desirable. The problem of patch potentials  is regarded as a major obstacle for present Casimir experiments,  and dedicated techniques based on Kelvin probe force microscopy are being developed to achieve a direct observation of the patches with the necessary spatial resolution \cite{deccapatch}.

In a recent paper \cite{hide}, the author proposed a setup based on ferromagnetic Ni  surfaces  for observing the thermal Casimir force, that 
is in principle immune from the problem of patch potentials.  The setup of  \cite{hide} consists of two aligned sinusoidally corrugated Ni surfaces, one of which is “hidden” by a thin opaque gold layer with a {\it  flat} exposed surface. 
The scheme of measurement is based on the observation of the phase-dependent modulation of the Casimir force, as one of the two corrugations is laterally displaced, relative to the other.  It was shown in \cite{hide}  that even for submicron separations between the plates the amplitude of the modulation predicted by the Drude model is several orders of magnitudes larger  than the modulation predicted by the plasma model. 
Preliminary data of an experiment   based on the scheme of \cite{hide}  can be found in the slides
of a talk   by R.S. Decca \cite{experdecca}.

The setup described in \cite{hide}  relies of the peculiar properties of the Casimir force between ferromagnetic surfaces. Ferromagnetic materials have been utilized in Casimir experiments only very recently  \cite{bani2}. In view of the  controversies surrounding the current debate on the thermal Casimir force, we consider it desirable to devise an apparatus that, while sharing the virtues of the setup \cite{hide}, is however based on nonmagnetic dielectric materials that have been routinely used in Casimir experiments for many years, and are therefore better understood. This is the aim of the present Letter.

\begin{figure}
\includegraphics{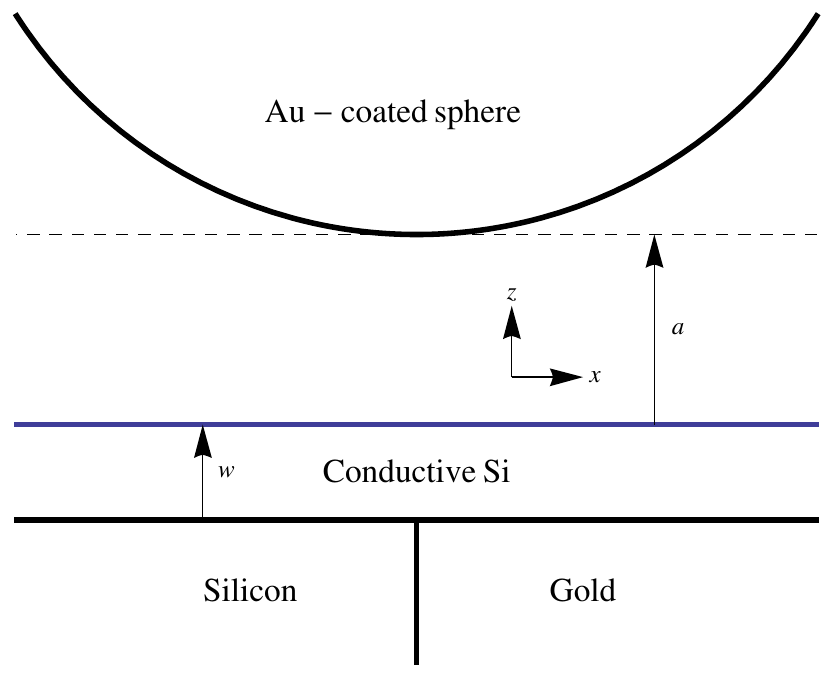}
\caption{\label{setup}   The setup consists of a Au-coated sphere   above a  planar slab divided in two regions,  respectively made of  high-resistivity (dielectric) Si and Au. Both the Si  and the Au regions are covered  with a semi-transparent plane-parallel conducting over-layer of uniform thickness $w$ made of low-resistivity Si. }
\end{figure}

The setup, schematically shown in Fig.\ref{setup},  consists of a gold sphere of radius $R$ above a planar slab divided in two regions,  made of  high-resistivity (dielectric) Si and Au respectively. 
The key feature of the apparatus is the {\it semi-transparent} \footnote{The over-layer needs be semitransparent in the IR, which is the important spectral region for the thermal force.} plane parallel { \it conducting} over-layer of thickness $w$, covering both the Si   and the Au regions. 
For any fixed sphere-plate separation $a$, we consider measuring  the {\it difference} 
\be\Delta F(a)=F_{\rm Si}(a)-F_{\rm Au}(a)\ee among  the values $F_{\rm Si}$ and $F_{\rm Au}$  of  the (normal) Casimir force that obtain when the tip of the sphere is respectively above a point $q$   deep in the Si region, and a point $p$   deep in the Au region. The great advantage of this differential mesurement over an absolute force measurement is that the  detrimental (mean) electrostatic force caused by patches on the exposed surfaces of the plates is automatically subtracted out from $\Delta F$, provided of course that the surface of the over-layer  has identical patch structure above the Si and Au regions of the plate.   

The idea of using a conducting over-layer to screen out electrostatic forces is not new per se, as it has been used before by the  IUPUI group in isoelectronic differential force measurements, searching for non-newtonian gravitational forces in the sub-micron region \cite{deccaiso,decca7}. A detailed study of the 
effect of random spatial variations of patch forces on the sensitivity of these experiments can be found  in \cite{speake}.   There is however a crucial difference with our scheme. In order to observe small differences between the gravitational forces on the Si and Au sectors, the Au overlayer of \cite{deccaiso,decca7} was designed to be opaque, to screen out altogether the otherwise dominant electrostatic and Casimir forces.  Its thickness ($w=150$ nm) was therefore chosen to be {\it  larger} than the plasma length ($\lambda_p$=130 nm for Au). In our setup, instead, we do want to observe the differential Casimir interaction of the Au sphere with the Au and Si regions of the plate. The conducting over-layer should thus screen the unwanted electrostatic component of the force, without affecting too much  the Casimir interaction with the underlying Au-Si structure. To do that, our over-layer needs to be semi-transparent and therefore its thickness has to be {\it much smaller} than the plasma length $\lambda_p$. This condition guided us to consider   B-doped {\it low-resistivity} Si as a candidate,   because its large plasma length ($\lambda_p=2.7 \mu$m)  ensures a good  transparency also for relatively large thicknesses $w$ (we take $w=100$ nm).  B-doped Si has been used alredy in Casimir experiments (see the second of Refs. \cite{semic}). Below, we shall use the symbol ${\rm Si_c}$ to denote conductive silicon, while the symbol Si shall denote dielectric silicon.

We now turn to the computation of the  Casimir force difference $\Delta F$. The framework for computing the Casimir force between real materials  is provided by Lifshitz theory \cite{lifs}. In its original version the  plates are planar dielectric surfaces ($\mu=1$)  fully described by the respective frequency-dependent  (complex) dynamical permittivity $\epsilon(\omega)$. 
 We  assume, as it is usually the case in Casimir experiments, that the sphere radius $R$ is much larger than the  separation $a$, and then it is possible to   estimate the Casimir force by the Proximity Force Approximation (PFA).  The PFA has been widely used to interpret most  Casimir experiments \cite{book2}  (see  \cite{kruger} for more applications of the Proximity Approximation). 
Recently, it has been shown that the PFA represents  the leading term in a gradient expansion of the Casimir force, in powers of the slopes of the bounding surfaces \cite{fosco2,bimonte3,bimonte4}. The gradient expansion shows that the PFA is asymptotically exact in the zero-curvature limit, and with its help it is now possible  to estimate the error introduced by the PFA.  
Besides $R \gg a$, two further conditions are required to ensure the validity of the PFA in our setup.  To be definite, we let $(x,y,z)$ be cartesian coordinates  such that $(x,y)$   span the exposed surface    of the ${\rm Si_c}$ over-layer, placed at $z=0$, while the sphere tip is at $z=a$. We imagine that the $x<0$ region of the slab is made of Si, while its $x>0$ region is made of Au.  
To ensure that we can neglect the effect of the sharp boundary between the Si and the Au regions, we 
assume that the horizontal distances of the points $p$ and $q$  from the Au-Si boundary are both much larger than the typical interaction radius $\rho=\sqrt{a R}$ of the circular region of the plate that contributes significantly to the Casimir interaction: $s \gg \rho$.  The force $F_{\rm Si}$ ($F_{\rm Au}$) is then undistinguishable from the force ${\tilde F}_{\rm Si}$ (${\tilde F}_{\rm Au}$) between the Au sphere and a plane-parallel two-layers slab consisting of a  ${\rm Si_c}$  layer of thickness $w$ on top of an infinite dielectric Si (Au) planar slab.  
Under these conditions, we  have for $\Delta F$:
$$
\Delta F \simeq {\tilde F}_{\rm Si}- {\tilde F}_{\rm Au} \simeq {F}_{\rm AuSi_c Si}^{(\rm PFA)}-{F}_{\rm AuSi_c Au}^{(\rm PFA)}
$$
\be
=2 \pi R ({ {\cal F}}_{\rm Au Si_c Si}- { {\cal F}}_{\rm AuSi_c Au})\;,\label{forpfa}
\ee 
where ${ { F}}_{\rm AuSi_c Si}^{(\rm PFA)}$ and ${ { F}}_{\rm AuSi_c Au}^{(\rm PFA)}$ denote the PFA expressions for the Casimir force between a Au sphere and a two-layer slab consisting of  a ${\rm Si_c}$ layer on top of either a Si or a Au subtrate, respectively, while  ${ {\cal F}}_{\rm AuSi_c Si} $ and ${\tilde {\cal F}}_{\rm AuSi_c Au} $ denote the unit-area Casimir free energies  for the corresponding plane-parallel systems in which the Au sphere is replaced by a Au planar slab. In the last passage of Eq. (\ref{forpfa}) we used the well-known PFA formula for the Casimir free energy of a sphere-plate system $
F^{(\rm PFA)}_{\rm sp-pl}(a) = 2 \pi R {\cal F}(a) $.  
The Casimir free-energy  ${\cal F}$ (per unit area) between two magnetodieletric possibly layered plane-parallel   slabs $S_{j}$, $j=1,2$ at distance $a$ in vacuum  can be represented by the formula: 
$$
{\cal F}(T,a)=\frac{k_B T}{2 \pi}\sum_{l=0}^{\infty}\left(1-\frac{1}{2}\delta_{l0}\right)\int_0^{\infty} d k_{\perp} k_{\perp}  
$$
\be
\times \; \sum_{\alpha={\rm TE,TM}} \log \left[1- {e^{-2 a q_l}}{R^{(1)}_{\alpha}({\rm i} \xi_l,k_{\perp})\;R^{(2)}_{\alpha}({\rm i} \xi_l,k_{\perp})} \right]\;,\label{lifs}
\ee
where $k_B$ is Boltzmann constant, $\xi_l=2 \pi l k_B T/\hbar$ are the (imaginary) Matsubara frequencies, $k_{\perp}$ is the modulus of the in-plane wave-vector, $q_l=\sqrt{\xi_l^2/c^2+k_{\perp}^2}$, and $R^{(j)}_{\alpha}({\rm i} \xi_l,k_{\perp})$ is the reflection coefficient of slab $j$ for polarization $\alpha$.  The Casimir free energy ${\cal F}_{\rm AuSi_c Si}$
for a planar Au-${\rm Si_c}$-Si system, can be  obtained from  Eq. (\ref{lifs}) by substituting  $R^{(1)}_{\alpha}$ by the Fresnel reflection coefficient $r_{\alpha}^{(0{\rm Au})}$ of a Au slab (given in Eqs.(\ref{freTE}) and (\ref{freTM}) below, with $a=0$, $b=$Au),  and $R^{(2)}_{\alpha}$ by the reflection coefficient  $R_{\alpha}^{(0{\rm Si_c Si})}$ of a two-layer planar slab consisting of a layer of thickness $w$ of conductive ${\rm Si_c}$ on a dielectric Si substrate.  The latter reflection coefficient has the expression:
\be  
R_{\alpha}^{(0{\rm Si_c Si})}({\rm i} \xi_l,k_{\perp};w)=\frac{r_{\alpha}^{(0{\rm Si_c})}+e^{-2\,w\, k_l^{({\rm Si_c})}}\,r_{\alpha}^{({\rm Si_c Si})}}{1+e^{-2\,w\, k_l^{({\rm Si_c})}}\,r_{\alpha}^{(0{\rm Si_c})}\,r_{\alpha}^{({\rm Si_c Si})}}\;.
\ee   
Here $r^{(ab)}_{\alpha}$ are the Fresnel reflection coefficients for a planar dielectric (we set $\mu=1$ for all materials) interface separating medium a from medium b: 
\be
r^{(ab)}_{\rm TE}=\frac{  \,k_l^{(a)}-  \,k_l^{(b)}}{ k_l^{(a)}+ \,k_l^{(b)}}\;,\label{freTE}
\ee
\be
r^{(ab)}_{\rm TM}=\frac{\epsilon_{b} ({\rm i} \xi_l) \,k_l^{(a)}-\epsilon_{a}({\rm i} \xi_l) \,k_l^{(b)}}{\epsilon_{b}({\rm i} \xi_l) \,k_l^{(a)}+\epsilon_{a}({\rm i} \xi_l) \,k_l^{(b)}}\;,\label{freTM}
\ee
where
$ k_l^{(a)}=\sqrt{\epsilon_a({\rm i} \xi_l)  \xi_l^2/c^2+k_{\perp}^2}\;$, 
 $\epsilon_a$  denotes the electric   permittivity of medium $a$, and we define $\epsilon_0=1$. The Casimir free energy ${\cal F}_{\rm AuSi_c Au}$ for a planar Au-${\rm Si_c}$-Au system
can be obtained by replacing Au to Si in the above formulae.

The values $\epsilon_{\rm a}({\rm i} \xi_l)$, (a=Au, Si)  of the permittivities of Au and dielectric Si were computed by means of the standard Kramers-Kronig disperion relation \cite{book2}
on the basis of the tabulated optical data quoted in \cite{palik}. The data for Au were extrapolated towards low frequencies by the Drude model, with Drude parameters  $\omega_{\rm Au}=9 \,{\rm eV}/\hbar$,  $\gamma_{\rm Au}=0.035\, {\rm eV}/\hbar$. For the permittivity of conductive Si  we used the formula
$\epsilon_{\rm Si_c}({\rm i} \xi)=\epsilon_{\rm Si}({\rm i} \xi)+\omega_{{\rm Si_c}}^2/\xi(\xi + \gamma_{\rm Si_c})$, with plasma frequency $\omega_{{\rm Si_c}}=7 \times 10^{14}$ rad/s 
and relaxation frequency $ \gamma_{\rm Si_c}=1.5 \times 10^{14}$ rad/sec  (see Ref. \cite{book2}, pag. 588). When using the plasma prescription to compute $\Delta F$, the TE $l=0$ term in Eq. (\ref{lifs}) is estimated by setting $\gamma_{\rm Si_c}=\gamma_{\rm Au_c}=0$. 
 
\begin{figure}
\includegraphics{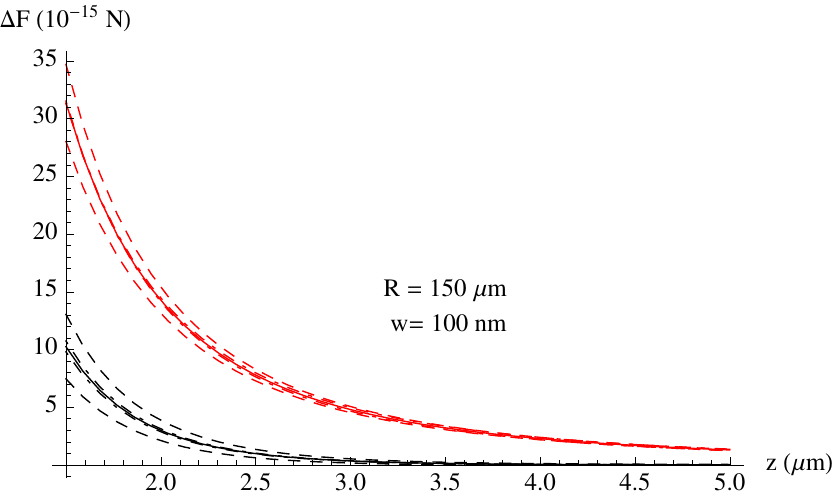}
\caption{\label{fig2}  The  force difference $\Delta F$ (in fN) versus separation $a$ (in microns). The red and grey lines correspond to  the plasma and Drude prescriptions, respectively. The dashed (dot-dashed)  lines show the effect of a 20 nm step (over-layer thickness difference $\delta w=20$ nm) across the Au-Si regions.}
\end{figure}

\begin{figure}
\includegraphics{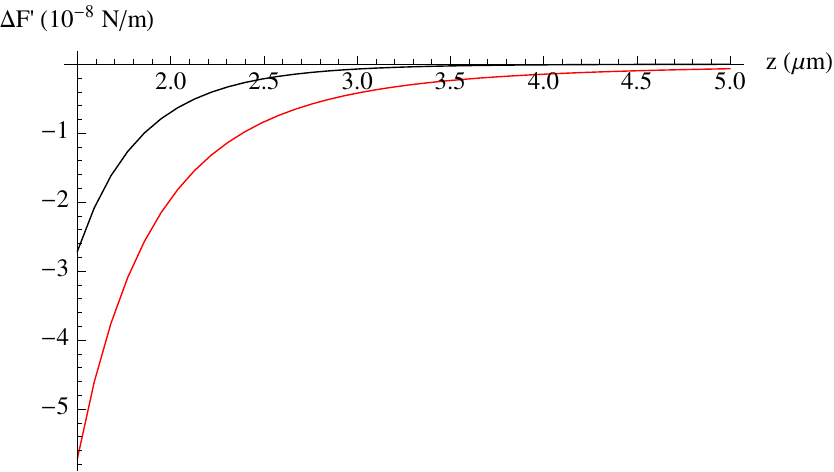}
\caption{\label{fig3new}  Difference of the Casimir force-gradient   $\Delta F '$ (in units of $10^{-8}$ N/m) versus separation (in $\mu$m). The red and grey lines  correspond to the plasma and Drude prescriptions, respectively. All parameters are as in Fig. \ref{fig2}.}
\end{figure}
In Fig. \ref{fig2} we plot   $\Delta F $   (in fN) versus separation $a$ (in $\mu$m) for a sphere of radius $R=150\; \mu$m and for a thickness $w=100$ nm of the ${\rm Si_c}$ over-layer. The red and the grey curves correspond the plasma and  Drude prescriptions, respectively. As we see, the plasma and the Drude models predict widely different magnitudes for $\Delta F$. For example, for $a=3$ $\mu$m, $\Delta F_{\rm plasma}$ is fourteen times larger than $\Delta F_{\rm Drude}$, while for $a=4$ $\mu$m they differ by a factor around fifty.   The signal is robust against imperfections in the sample geometry. This can be seen from Fig. \ref{fig2}, where the dashed (dot-dashed) lines  show the effect of a 20 nm step (over-layer thickness difference $\delta w$=20 nm) across the Au-Si regions. Roughness has a negligible effect as well. The  recent  isoelectronic differential experiment at IUPUI \cite{decca7}, using a  Au coated sphere  of radius $R \simeq 150 \mu$m, achieved a sensitivity better than 0.3 fN in force differrences,  independent of the separation $a$ in the range from 200  nm to 1 $\mu$m.   With  this level of  sensitivity, it would be possible to accurately measure $\Delta F$   up to separations of several $\mu$m, and then it should be possible to easily discriminate between the plasma and the Drude prescriptions.  In recent years very high sensitivities have been demonstrated also in dynamical experiments \cite{decca4,decca5,decca6,chang2} which measure the gradient of the Casimir force $F'$.   It is thus useful to plot the change in the gradient of the Casimir force $\Delta F ' $ for our apparatus. In  Fig. \ref{fig3new} $\Delta F ' $ (in units of $10^{-8}$ N/m)   is plot versus separation $a$ (in $\mu$m) for the same values of $R$ and $w$ as in Fig \ref{fig2}.

We have described a  Casimir apparatus for observing the thermal Casimir force   in the $\mu$m range.
It involves a differential measurement of the Casimir force between   a Au-coated sphere  and a  planar slab  consisting of two regions, respectively made of  high-resistivity (dielectric) Si and  Au.
The key feature of the setup is  a {\it semi-transparent} plane parallel conducting over-layer covering both the Si  and the Au regions, whose purpose is to filter out  unwanted electrostatic forces caused by potential patches on the plates surfaces, that plague  present large distance Casimir experiments.  Metallic {\it opaque}  over-layers have been successfully used to electrically screen test masses in recent searches of non-newtonian gravitational forces based on  isoelectronic force-difference measurements \cite{deccaiso, decca7}.   The 0.3 fN sensitivity reached by  these experiments    makes us expect that with our apparatus it should be possible to observe the thermal Casimir force up to separations of several $\mu$m, and to clearly discriminate between the Drude and the plasma prescriptions for the thermal force.
  
The main experimental challenge posed by our setup is probably the realization of a semi-transparent conductive overlayer with identical patch structure over the Au-Si regions of the plate, which is indispensible for nullifying the mean patch electrostatic force in the Casimir force difference $\Delta F$. This goal was achieved with opaque Au over-layers in the experiments \cite{deccaiso,decca7}. The choice of conducting Si made in this work may not be  optimal. 
The identification of the best material and deposition technique for the over-layer  is a task that can only be addressed by  experiment.

\end{document}